\documentclass[journal]{IEEEtran}
\usepackage{cite}

\ifCLASSINFOpdf
  \usepackage[pdftex]{graphicx}
\else
\fi
\usepackage{amsmath}
\usepackage{amssymb}
\usepackage{algorithmic}

\usepackage{array}
\begin{document}
%
\title{Unsupervised Low-light Image Enhancement with Decoupled Networks}
%
%
%

\author{Wei Xiong,
        Ding Liu,
        Xiaohui Shen,
        Chen Fang,
        and Jiebo Luo
\thanks{Wei Xiong and Jiebo Luo are with University of Rochester. (Email: wei.xiong@rochester.edu; jluo@cs.rochester.edu)}
\thanks{Ding Liu and Xiaohui Shen are with ByteDance Inc. (Email: liuding@bytedance.com; shenxiaohui@bytedance.com)}
\thanks{Chen Fang is with Tencent Inc. (Email: fangchen1988@gmail.com)}
}

%
%

\markboth{Journal of \LaTeX\ Class Files,~Vol.~14, No.~8, August~2015}%
{Shell \MakeLowercase{\textit{et al.}}: Bare Demo of IEEEtran.cls for IEEE Journals}
%



\maketitle

\begin{abstract}
In this paper, we tackle the problem of enhancing real-world low-light images with significant noise in an unsupervised fashion. Conventional unsupervised learning-based approaches usually tackle the low-light image enhancement problem using an image-to-image translation model. They focus primarily on illumination or contrast enhancement but fail to suppress the noise that ubiquitously exists in images taken under real-world low-light conditions. To address this issue, we explicitly decouple this task into two sub-tasks: illumination enhancement and noise suppression. We propose to learn a two-stage GAN-based framework to enhance the real-world low-light images in a fully unsupervised fashion. To facilitate the unsupervised training of our model, we construct samples with pseudo labels. Furthermore, we propose an adaptive content loss to suppress real image noise in different regions based on illumination intensity. In addition to conventional benchmark datasets, a new unpaired low-light image enhancement dataset is built and used to thoroughly evaluate the performance of our model. Extensive experiments show that our proposed method outperforms the state-of-the-art unsupervised image enhancement methods in terms of both illumination enhancement and noise reduction.
\end{abstract}


%
\IEEEpeerreviewmaketitle

\section{Introduction}
%
%
%
%
\IEEEPARstart{R}eal-world low-light image enhancement \cite{lee2007efficient,ren2018lecarm,zhao2021retinexdip,li2021low} is challenging since images captured under low-light conditions usually exhibit low illumination and contain heavy noise. Enhancing these images requires adjusting illumination, contrast, color, as well as suppressing the noise while preserving the details simultaneously. 
Traditional methods for this task primarily focus on adjusting contrast via a fixed tone-mapping \cite{mantiuk2008display}, resulting in limited performance on challenging cases. Recently, learning-based methods have been utilized to learn content-aware illumination enhancement from data with deep neural networks \cite{gharbi2017deep,wang2019underexposed,xu2020learning,guo2020zero,moran2020deeplpf,yang2020fidelity}. Despite promising  performance, many of them heavily rely on low-light and normal-light image pairs, which are expensive or even impossible to obtain in real-world scenarios. One alternative way to cheaply generate such training pairs is to synthesize a low-light image from its counterpart captured under a normal light condition \cite{lore2017llnet}. However, due to the significant signal distribution gap between synthesized dark images and ones taken under real-world low-light conditions, models trained on synthesized image pairs usually fail to generalize well in realistic scenarios \cite{guo2019toward,martin2017rethinking}.

Recently, several unsupervised deep learning-based methods have been developed for image enhancement to eliminate the reliance on paired data. As a general-purpose method, unsupervised image-to-image translation or transformation models \cite{isola2017image,liu2017unsupervised,zhu2017unpaired,xiong2020fine} such as CycleGAN \cite{zhu2017unpaired}, and UNIT \cite{liu2017unsupervised} can be applied to image enhancement. These methods adopt generative adversarial networks (GANs) \cite{goodfellow2014generative} to encourage the distribution of the generated images to be close to that of the target images without paired supervision. Recently, the GAN-based models have been specially designed to address the task of illumination enhancement \cite{lore2017llnet,chen2018deep,jiang2019enlightengan,guo2020zero,yang2020fidelity}. These unsupervised learning approaches can generate images with better illumination and color in several cases. However, they have common limitations in the real-world low-light image enhancement task in two aspects (as we will show in our experiments later):
1) the contrast and illumination of enhanced images can be unsatisfactory and usually suffer from color distortion and inconsistency. The bright regions of a dark image can be  over-exposed. Moreover, continuous regions may exhibit a sharp color or brightness inconsistency due to the unstable training of the unsupervised models. 
2) when applied to low-light images with heavy noise, models with a single image-to-image mapping network primarily address illumination enhancement but usually fail on noise suppression. \textit{This suggests that a single network may be inadequate to model both the complex illumination patterns and real-world noise patterns.} Although a few prior works have been developed to remove noise after the illumination enhancement as a post-processing step, they either use traditional methods such as BM3D \cite{dabov2007image} that requires a given noise level as an input, or use learning-based denoising methods that are originally designed for synthesized noise such as additive white Gaussian noise \cite{lehtinen2018noise2noise,krull2019noise2void,batson2019noise2self}.


To address these issues, we propose to explicitly decouple the whole unsupervised enhancement task into two sub-tasks: 1) illumination enhancement and 2) noise suppression. We propose a two-stage framework to handle each sub-task separately. 
Specifically, in Stage I, a Retinex-based deep network is trained under a GAN framework in an unsupervised manner to enhance the illumination of low-light images while preserving the contextual details. Instead of predicting the final enhanced image directly, we predict the illumination map of the target image first then use the predicted illumination to generate the enhanced image. It has been suggested by recent low-light image enhancement works \cite{wang2019underexposed} that the illumination maps for natural images usually have simple forms. Therefore, the Retinex-based illumination modeling can benefit the generalization ability of the generator. To tackle the inconsistency problem in color and brightness, we introduce a pyramid module to enlarge the receptive field of the generator.

In Stage II, we propose a guided unsupervised denoising model based on GANs. Our model is adaptively guided by the illumination conditions of the original low-light image and the enhanced image from Stage I. Inspired by the success of pseudo labeling methods in semi-supervised learning \cite{lee2013pseudo,berthelot2019mixmatch}, 
in this work, we propose to construct \textit{pseudo triples}, i.e., a pseudo low-light image, a pseudo image after illumination enhancement, and a real noise-free normal-light image with the same content at each training iteration, to facilitate the unpaired training for image denoising.
We find this technique is crucial to the success of unsupervised noise modeling. 
We design an adaptive content loss for the \textit{pseudo triples} to preserve the illumination and color of the input image. 

It is noteworthy that a recent work, GCBD~\cite{chen2018image}, also adopts a GAN-based denoising framework that can be learned without paired data. However, our Stage II model significantly differs from this work in the following aspects. First, GCBD limits noise estimation in smooth regions only before learning the GAN-based denoiser, while we jointly perform noise modeling and noise removal in a single GAN framework. Second, GCBD does not consider the influence of illumination. In contrast, our denoising model is explicitly guided by the illumination conditions of the input images to  handle illumination-correlated noise adaptively. Third, we propose an adaptive content loss using \textit{pseudo triples} to remove real image noise based on the illumination intensity, while GCBD may produce severe color distorted results without such a constraint, as demonstrated by our experiments later. 

We evaluate the performance of our proposed approach over the LOw-Light (LOL) dataset \cite{wei2018deep} and an unpaired enhancement dataset from \cite{jiang2019enlightengan}.
To further demonstrate the effectiveness of our method, we contribute an Unpaired Real-world Low-light image enhancement dataset (URL) for evaluation. Our dataset is composed of 1) low-light images captured under real-world low-light conditions with varying levels of noise, and 2) normal-light images collected from existing data galleries, which consist of diverse scenes ranging from outdoor scenes to indoor pictures. We compare our method with the state-of-the-art unsupervised learning-based enhancement methods on these datasets. Extensive experiments show that our method outperforms other methods in terms of both illumination enhancement and noise suppression.

Our primary contributions are: 
\begin{itemize}
    \item We propose a decoupled framework for unsupervised low-light image enhancement; 
    \item We propose an illumination guided unsupervised denoising model. To facilitate unsupervised training of our denoising model, we construct \textit{pseudo triples} and propose an adaptive content loss to denoise regions guided by both the original lighting condition and enhanced illumination;
    \item We build an unpaired low-light image enhancement dataset containing varying noise and good diversity as an important complement to the existing low-light enhancement datasets. 
\end{itemize}

\section{Related Work}
\subsection{Image Enhancement.}
Traditional image enhancement methods are primarily built upon histogram equalization (HE) \cite{pizer1987adaptive} or Retinex theory \cite{land1977retinex,jobson1997multiscale}. 
HE-based methods aim to adjust the histogram of pixel intensities to obtain an image with better contrast \cite{pizer1987adaptive}. Retinex-based methods assume that an image is the composition of illumination map and reflectance, and thus low-light images can be restored by estimating the illumination map and reflectance map \cite{jobson1997multiscale}. 
Recently, learning-based methods have been proposed to learn the illumination enhancement from data \cite{fu2016weighted,guo2016lime,li2018structure}. Later deep neural networks have been used and achieved promising results \cite{lore2017llnet,gharbi2017deep,wei2018deep,chen2018learning}.  A more recent work \cite{moran2020deeplpf} proposes the DeepLPF model to enhance images by learning different types of filters. \cite{xu2020learning} \textit{et al.} propose a frequency-based decomposition-and-enhancement model for low-light image enhancement. However, the learning of these models heavily rely on image and reference pairs. 

Due to the difficulty of acquiring paired data in real-world scenarios,
several weakly supervised and unsupervised enhancement approaches have been proposed, such a WESPE \cite{ignatov2018wespe}, Deep Photo Enhancer \cite{chen2018deep}, and EnlightenGAN \cite{jiang2019enlightengan}. 
Ignatov et al. propose a transitive GAN-based enhancement model that can be learned without paired data \cite{ignatov2018wespe}. Chen et al. propose an unpaired model based on 2-way GANs to enhance images \cite{chen2018deep}. Jiang et al. further propose EnlightenGAN which is specifically designed for illumination enhancement of low-light images \cite{jiang2019enlightengan}.
There are also general-purpose unsupervised image translation models that can be used for image enhancement, such as CycleGAN \cite{zhu2017unpaired}, UNIT \cite{liu2017unsupervised} and ADN \cite{adn2019_miccai}. 
A primary limitation of existing enhancement approaches is that they mainly perform illumination enhancement and fail to address noise suppression, such as CycleGAN \cite{zhu2017unpaired} and EnlightenGAN \cite{jiang2019enlightengan}. In contrast, we address both illumination enhancement and noise suppression simultaneously. 

It is worth noting that a recent work named Zero-DCE \cite{guo2020zero} also tackles the image enhancement problem in an unsupervised fashion. They learn a curve estimation model with deep networks and use it for pixel-wise dynamic range adjustment of the input image. However, their work primarily focuses on contrast enhancement without paying much attention to noise removal.  

\subsection{Real-world Image Denoising.}
There have been a number of works for image denoising, including conventional methods such as BM3D \cite{dabov2007image} and Non-local means \cite{buades2005non}, and deep learning-based models such as DnCNN \cite{zhang2017beyond}, Residual Dense Networks \cite{zhang2020residual} and Non-local Recurrent Networks \cite{liu2018non}.  However, most of the models are limited to synthetic noise removal and are difficult to generalize to real-world noise removal. Recently, real-world blind denoising models have been proposed to learn a blind denoiser from real-world paired data \cite{xu2017multi,guo2019toward,kim2019grdn,lebrun2015multiscale,zhang2018ffdnet,xu2018trilateral}. 
Xu et al. \cite{xu2017multi} design a multi-channel weighted nuclear norm minimization model to use channel redundancy.
Guo et al. propose CBDNet \cite{guo2019toward} to directly learn a blind denoiser from real-world paired data. Kim et al. leverage a GAN based deep network for real-world noise modeling \cite{kim2019grdn}. Other approaches \cite{lebrun2015multiscale,zhang2018ffdnet,xu2018trilateral} also show promising results.

Most learning-based denoising models need to be trained with paired data, which is expensive to obtain for real-world noise removal tasks. Recently, several unsupervised denoising methods have been devised, including self-supervised learning approaches, such as Noise2Noise \cite{lehtinen2018noise2noise} and Noise2Void \cite{krull2019noise2void}, as well as unpaired training approaches \cite{chen2018image,yan2019unsupervised}. 
Our Stage II model is also an unsupervised denoising approach. Unlike previous methods, we capture the real-world noise pattern with the explicit guidance of image illumination conditions and denoise the images with \textit{Pseudo Labeling} technique and an adaptive content loss, which prove to be crucial to the success of real-world image denoising. 


\begin{figure*}[t]
    \centering
    \includegraphics[width=\textwidth]{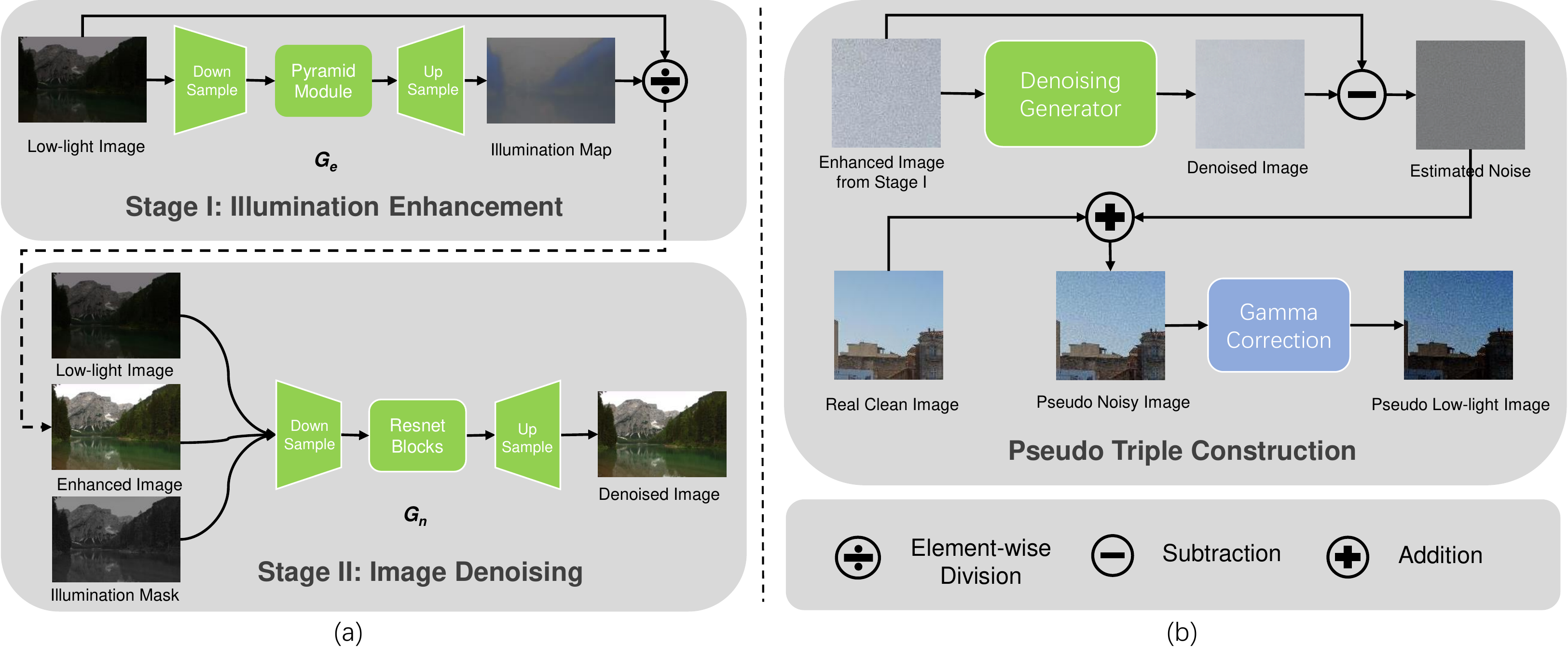}
    \caption{(a): An overview of our proposed two-stage decoupled networks. In Stage I, given a low-light image, we learn a deep network to predict an illumination map, then use the Retinex theory to obtain the illumination-enhanced image. Then in Stage II, we input the original low-light image, the enhanced image from Stage I and the illumination mask to generate an image with reduced noise and better details. The illumination mask indicates how much the illumination is improved in Stage I. (b): An illustration of pseudo triple construction. We first estimate a noise image from our noisy training data. Then we randomly fetch a clean normal light image and add the estimated noise to the clean image to obtain a pseudo noisy image. Next, we perform Gamma correction to obtain a pseudo low-light counterpart of the noisy image. In this way, we obtain a pseudo triplet  \{pseudo low-light image, pseudo noisy image, real clean image\} as additional training data to train our denoising model.}
    \label{fig:framework}
\end{figure*}

\section{Our Approach}
As shown in Fig. \ref{fig:framework} (a), our approach for real-world low-light image enhancement consists of two stages. In Stage I, we perform illumination enhancement on the real-world low-light images while preserving contextual details. 
In Stage II, we propose an unsupervised learning-based denoising model to suppress the noise in the output image from Stage I and enhance the contextual details. 

\subsection{Stage I: Illumination Enhancement}
Given a noisy low-light image $I_l$, our goal in this stage is to learn a model $G_e$ to generate an enhanced image $I_e$ with proper illumination, color, as well as realistic content details. Conventional learning-based models such as CycleGAN and EnlightenGAN usually adopt a U-net \cite{ronneberger2015u} like architecture to predict the enhanced image from the low-light input image directly. 
However, under the unsupervised learning scheme, directly applying such an  architecture is easy to produce results with unstable illumination, such as color distortion or inconsistency \cite{jiang2019enlightengan}. Recent work \cite{wang2019underexposed} on low-light image enhancement suggests that illumination maps for natural images usually have relatively simple forms, thus  using Retinex-based illumination modeling can facilitate the learning process, leading to an enhancer with better generalization ability. Inspired by this, 
we adopt a Retinex-based model to enhance the low-light images. Based on the Retinex theory, an image $I$ can be modeled as $I = S \circ R$, where $S$ is the illumination, $\circ$ denotes element-wise multiplication, and $R$ is the reflectance. Similar to \cite{wei2018deep,wang2019underexposed}, we regard the reflectance as a well-exposed image $I_e$, then we have $I_l = S \circ I_e$. In reverse, the enhanced image $I_e$ can be recovered from the low-light image $I_l$ given the predicted illumination map $S$ .
As shown in Fig. \ref{fig:framework}, we use the generator $G_e$ to estimate an illumination map $S=G_e(I_l)$ from low-light image $I_l$. Then we obtain the enhanced image 
\begin{equation}
    I_e = I_l / S,
\end{equation}
 where $/$ is element-wise division.

\textbf{Model Architecture.} As shown in Fig. \ref{fig:framework}, the input low-light image $I_l$ is passed to the encoder, then a pyramid module, and then decoded by the decoder into an illumination map $S$ with RGB channels. 
The pyramid module, inspired by PSPNet \cite{zhao2017pyramid}, is customized to enlarge the receptive field of our network. 
In this module, we down-sample the feature maps into features at multiple resolutions, namely, $1\times1$, $2\times2$, $4\times4$, $16\times16$. At each resolution, feature maps are followed by a convolution layer and a  ReLU~\cite{xu2015empirical}  layer. Then the transformed feature maps are upsampled and concatenated, then fused into the next convolution layer. The fused feature maps are then decoded by the decoder to generate the illumination map $S$. With the pyramid module, the receptive field of the model is enlarged, and the network can perceive the illumination information at different spatial levels, which is beneficial for reducing the color and contrast inconsistency. 

\textbf{Loss Functions.} 
In Stage I, the generator $G_e$ is trained with unpaired images. To achieve this goal, we adopt adversarial learning to encourage the distribution of the enhanced image $I_e$ to be close to that of the normal-light images. Specifically, we use two discriminators to distinguish the generated image from the real normal-light image. The global discriminator $D_g$ takes the whole image as the input, and outputs the realness of the image. The local discriminator $D_l$ takes random patches extracted from the image and outputs the realness of each patch. The global discriminator encourages the global appearance of the enhanced image to be similar to  a normal-light image, while the local discriminator ensures that the local context (shadow, local contrast, highlight, \textit{etc.}) can be as realistic as the real normal-light images. 

We adopt the LSGAN version of relativistic average GAN loss \cite{jolicoeur2018relativistic} for training the global discriminator. When updating the discriminator, we have:
\begin{equation}
\begin{split}
    L_D^{g} &= \mathbb{E}_{x_r \in \mathbb{P} } [(D_g(x_r) - \mathbb{E}_{x_f \in \mathbb{Q}}{D_g(x_f)} -1)^2] \\ 
    &+ \mathbb{E}_{x_f \in \mathbb{Q}}[(D_g(x_f) - \mathbb{E}_{x_r \in \mathbb{P}}D_g(x_r))^2]
\end{split}
\end{equation}

When updating the generator, we have:
\begin{equation}
\begin{split}
    L_G^{g} &= \mathbb{E}_{x_r \in \mathbb{P} } [(D_g(x_r) - \mathbb{E}_{x_f \in \mathbb{Q}}{D_g(x_f)})^2] \\
    &+ 
    \mathbb{E}_{x_f \in \mathbb{Q}}[(D_g(x_f) - \mathbb{E}_{x_r \in \mathbb{P}}D_g(x_r) -1)^2]
\end{split}
\end{equation}
where $\mathbb{P}$ and $\mathbb{Q}$ are the real image (the normal-light images) distribution and the generated image distribution, respectively. $x_r$ and $x_f$ are samples from distribution $\mathbb{P}$ and $\mathbb{Q}$, respectively. 

We adopt the original LSGAN loss for training the local discriminator. When updating the discriminator, we have:
\begin{equation}
    L_D^{l} = \mathbb{E}_{x_r \in \mathbb{P} } [(D_l(x_r) -1)^2] + 
    \mathbb{E}_{x_f \in \mathbb{Q}}[(D_l(x_f) )^2]
\end{equation}

When training the generator, we have:
\begin{equation}
    L_D^{l} = 
    \mathbb{E}_{x_f \in \mathbb{Q}}[(D_l(x_f) -1 )^2]
\end{equation}

Similar to \cite{jiang2019enlightengan}, we use a perceptual loss (computed on VGG features) between the output image and the input image to preserve content details from the input image. To prevent the output image from being as dark as the input image, we alleviate the influence of image brightness and force the network to focus on only the content preservation by using instance normalization on the VGG feature maps before performing the perceptual loss. Similar to the adversarial loss, we calculate perceptual loss on both the whole image and the image patches. We formulate the global perceptual loss $L_P^g$ and the local version $L_P^l$ as:
\begin{equation}
    L_P^{g} = \|\Phi(I_l) - \Phi(I_e)\|_2^2/N,
\end{equation}
\begin{equation}
    L_P^{l} = \|\Phi(I_l^p) - \Phi(I_e^p)\|_2^2/N,
\end{equation}
where $\Phi$ denotes to the VGG19 feature extractor, $N$ is the number of elements in the image, $I_l^p$ and $I_e^p$ are random patches extracted from $I_l$ and $I_e$, respectively. 

By minimizing both the adversarial losses and the perceptual losses, we are able to learn a good illumination predictor and produce results without color distortion and with better contextual details.
 
\textbf{Differences between Prior Work.}
The primary differences between our Stage I model and previous unsupervised low-light image enhancement method EnlightenGAN \cite{jiang2019enlightengan} lie in the theory we follow and the model architecture. First, EnlightenGAN uses an image-to-image translation model to map the input low-light image to the illumination-enhanced image directly. In contrast, our model is derived from the Retinex theory, which predicts the illumination map instead of the enhanced image. Such a model proves to be effective and has better generalization ability for image contrast enhancement. Second, we propose to use a pyramid module to connect the encoder and the decoder in our generator. Such a design enlarges the receptive field of our network and helps the model better perceive both the global and local illumination conditions.

\subsection{Stage II: Noise Suppression}

As shown in Fig. \ref{fig:framework}, our noise suppression model $G_n$ adopts the original low-light image $I_l$, the enhanced image (with noise) $I_e$ from Stage I, and an illumination mask $M$ as inputs to generate the final clean image $I_c$ with reduced noise as well as a good illumination condition, i.e., $I_c = G_n(I_l, I_e, M)$. $M$ serves as an indicator showing how much illumination is increased from low-light image $I_l$ to enhanced image $I_e$. We have $M = max(illu(I_e) - illu(I_l), 0)$, where $illu(\cdot)$ means extracting the illumination of an image. In our work, we directly use the gray-scale version of an image as its illumination map. With  $M$,  $I_l$, and  $I_e$ as inputs, our denoising model is explicitly guided by the illumination conditions.

\textbf{Model Architecture.}
Our denoising generator $G_n$ in Stage II adopts an encoder-decoder architecture, with several convolutional blocks followed by several Resnet blocks then decoded back to an image. We adopt a multi-scale discriminator \cite{liu2017unsupervised} to predict the realness of images at multiple resolutions. 
The detailed architecture of the networks can be found in the appendix section.

\textbf{Loss Functions.}
Since there is no ground-truth for the input low-light image during training, to learn a noise-free image from its noisy counterpart, we adopt an LSGAN-based adversarial loss to encourage the generated image to be as clean as the real-world clean normal-light images. Note that the discriminator needs not only to judge whether the illumination and color of the generated image are realistic enough but also needs to judge whether the generated image is clean without much noise. Training a single discriminator with clean images or synthesized images to do both tasks simultaneously is challenging. As our goal in this stage is noise suppression, when feeding the discriminator, we first perform an instance normalization on both the synthesized image and the normal-light clean image to reduce the influence of image illumination, color, and contrast. 

During training, we randomly match the denoised image $I_c$ to a normal-light clean image $J_c$. Our adversarial loss for training the discriminator $D_n$ is:

\begin{equation}
    L_D = \mathbb{E}_{J_c \in \mathbb{P} } [(D_n(J_c) -1)^2] +
    \mathbb{E}_{I_c \in \mathbb{Q}}[(D_n(I_c)^2] .
\end{equation}

The corresponding loss for updating the generator is $ L_G = \mathbb{E}_{I_c \in \mathbb{Q}}[(D_n(I_c) -1 )^2]$, 
where $\mathbb{P}$ and $\mathbb{Q}$ are the normal-light clean image distribution and the distribution of generated images in Stage II, respectively. 

Merely using the adversarial loss can cause color shifting problem, i.e., the color of the generated images can be easily distorted, since we only constrain the images after instance normalization to be similar to the normal-light images. As we have already obtained an image $I_e$ with a satisfactory contrast and color from Stage I, in Stage II, we only need to preserve the contrast and color of $I_e$. Therefore, we use a color loss to constrain the generated image $I_c$ to have the same color as $I_e$. 

Specifically, we first down-sample the images with average pooling to $I_c^{\downarrow}$ and $I_e^{\downarrow}$ to suppress the noise in $I_e$, then perform the color matching. We have 

$$L_{color} = \sum_p\angle((I_c^{\downarrow})_p,  (I_e^{\downarrow})_p) / N_n ,$$
where $p$ is the location of a pixel in the down-sampled image, $\angle(x,y)$ calculates the inner product between two 3-D vectors which are composed of RGB channels of a pixel location, $N_n$ is the number of pixels in the downsized image.

\textbf{Pseudo Labeling: Constructing Pseudo Triples for Unsupervised Learning. }
We propose a \textit{Pseudo Labeling} technique to facilitate the unsupervised training of the denoising model. As shown in Fig. \ref{fig:framework} (b), we first estimate the noise in image $I_e$ as $I_n = I_e - I_c$. Then given the randomly matched normal-light clean real image $J_c$, we can simulate a pseudo noisy image $J_e$ by adding the estimated noise to the clean image, i.e., $J_e = J_c + I_n$. We also use Gamma Correction \cite{gamma} to decrease the brightness of $J_e$, to obtain a corresponding pseudo low-light image $J_l=(J_e)^{\lambda}$, where $\lambda$ is estimated as $\lambda = \log \overline{I}_c / \log \overline{I}_l$. $\overline{I}_c$ and $\overline{I}_l$ are the average pixel values over all pixel locations of image $I_c$ and $I_l$, respectively. After these steps, we obtain a \textit{pseudo triple} $\mathcal{J} = \{J_l, J_e, J_c \}$,  
where $J_l$ is the constructed pseudo low-light image, $J_e$ is the pseudo enhanced image (with noise) and $J_c$ is the real clean image from our training set. 
Similarly, we construct the illumination mask for the pseudo triple as $M_J = max(illu(J_e)-illu(J_l) ,0)$. 
We can then predict a denoised image $J_g=G_n(J_l, J_e, M_J)$ from the constructed fake images, and use $J_c$ as the supervision to train $G_n$.

\textbf{Adaptive Content Loss with Pseudo Triples.} 
To train $G_n$, we adopt an adaptive content loss to constrain the generated image $J_g$ to be perceptually close to the clean normal-light image $J_c$. This is achieved by using both perceptual loss and L1 reconstruction loss on the pixel space between $J_g$ and $J_c$. As different regions may have different lighting conditions, regions with a significant brightness increase after the first stage may contain heavy noise, and regions without a large brightness increase may contain less noise. When imposing the reconstruction constraint, we encourage the network to focus more on dark regions where noise is usually heavier. We then formulate the adaptive content loss for the pseudo triples as:

\begin{equation}
    \begin{split}
    L_{con}^{adapt} &= \sum_l \|M_J^{(l)}\circ(\Phi_l(J_g) - \Phi_l(J_c))\|_2^2/N_l \\ 
    &+ \gamma_p \|M_J\circ(J_g - J_c)\|_1/N,
    \end{split}
\end{equation}
where $M_J^{(l)}$ is the downsized version of $M_J$ that matches the spatial size of the VGG features at the $l$-th VGG layer. $M_J$ serves as the weight mask for each pixel in the image.  $N$ is the number of elements in image $J_g$, $N_l$ is the number of elements in the feature maps of the $l$-th layer in VGG network. $\Phi_l(I)$ is the feature in the $l$-th layer of VGG given the input image $I$. $\gamma_p$ is the weight to balance the losses from the RGB image domain and the VGG feature domain. In our work, we choose the layers of ``relu1\_2'', ``relu2\_2'', ``relu3\_2'', ``relu4\_4'', ``relu5\_4'' to perform both low-level and high-level feature matching, and $\gamma_p$ is set as $10$. 
We do not use instance normalization on the VGG feature maps of the \textit{pseudo triple}, since we need to preserve the color and contrast.

\textbf{Interpretation of Pseudo Labeling.}
Note that at the early training stage, the estimated noise may contain clear structures of the objects in the input noisy image $I_e$. As a result, the constructed pseudo image $J_e$ may also contain object structures from $I_e$. Since our network is trained to remove noise, it will be difficult for the network to remove the high-frequency object structure patterns. Then the generated image $J_g$ may also contain object structures from $I_e$. By minimizing our proposed adaptive content loss, we are essentially encouraging the estimated noise to contain fewer object structures, i.e., encouraging the network to estimate the noise pattern more accurately. 

\textbf{Content Preserving Loss. }
To make sure that the denoised image $I_c$ preserves contextual details of the enhanced image $I_e$, we also impose a perceptual loss and a reconstruction loss between images $I_c$ and $I_e$. To reduce the influence of color and contrast, we perform instance normalization to the images before imposing the perceptual loss and reconstruction loss. The content loss is formulated as:

\begin{equation}
    L_{con} = \sum_l \|\Phi_l(I_e) - \Phi_l(I_c)\|_2^2/N_l + \gamma_c \|I_e - I_c\|_1/N ,
\end{equation}
where the layers and operations used are the same as $L_{con}^{adapt}$. 
$\gamma_c$ is a weight balance term similar to $\gamma_p$ and set as 10 in our work.
The total loss $L$ for training $G_n$ is a combination of all the losses.
\begin{equation}
    L = L_G + \lambda_{c} L_{color} + \lambda_C^{p} L_{con}^{adapt} + \lambda_C^{r} L_{con} ,
\end{equation}
and we empirically find that setting $\lambda_{c}, \lambda_C^{p}, \lambda_C^{r}$ as 10, 1, 1, respectively, yields the best result.

\begin{figure*}[t]
    \centering
    \includegraphics[width=\textwidth]{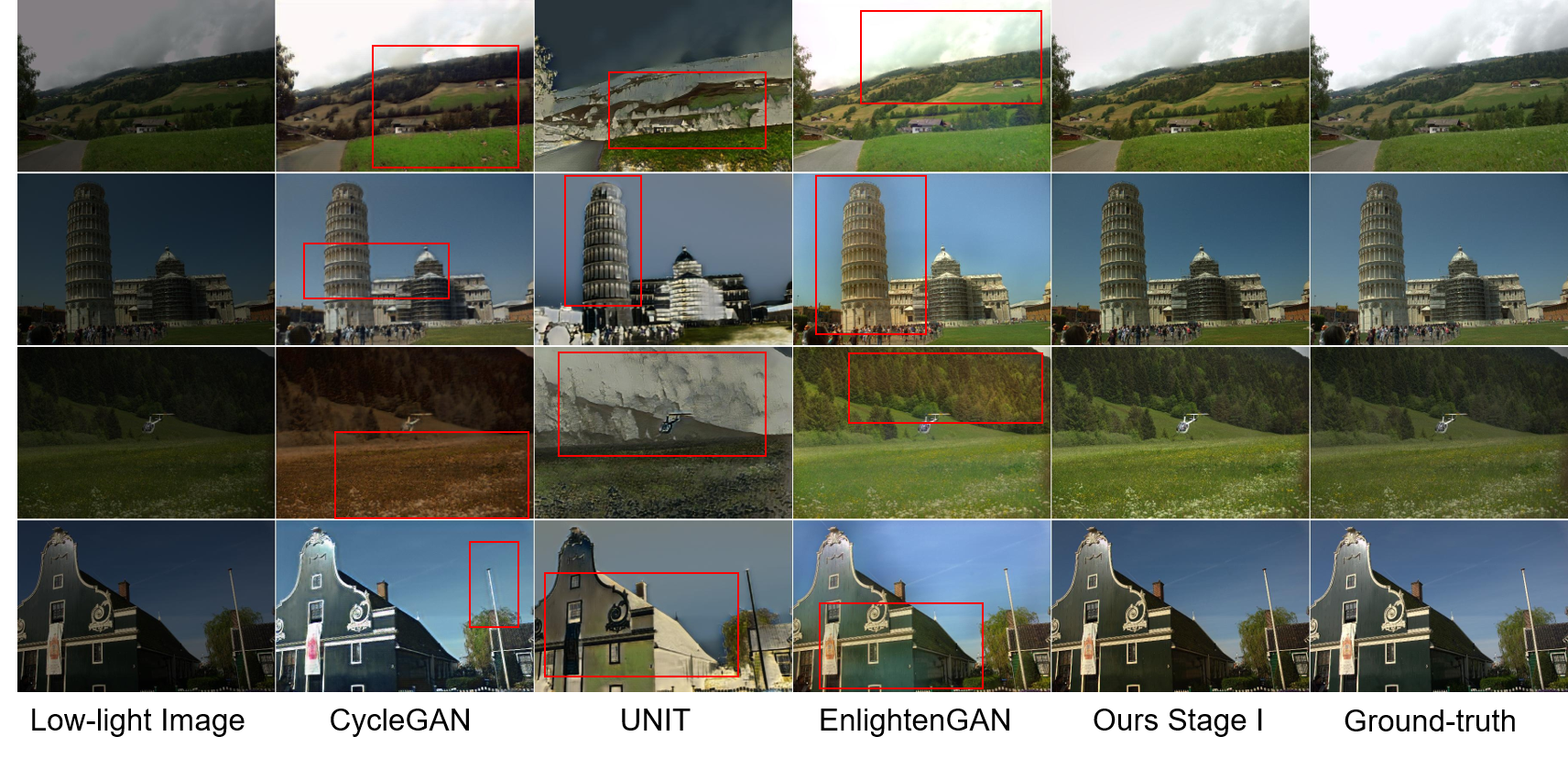}
    \caption{Illumination enhancement results on the Unpaired Enhancement Dataset from \cite{jiang2019enlightengan}. Please pay attention to the regions in the red boxes, where severe color distortion and contrast inconsistency occur.}
    \label{fig:contrast}
\end{figure*}

\begin{figure}[t]
    \centering
    \includegraphics[width=0.48\textwidth]{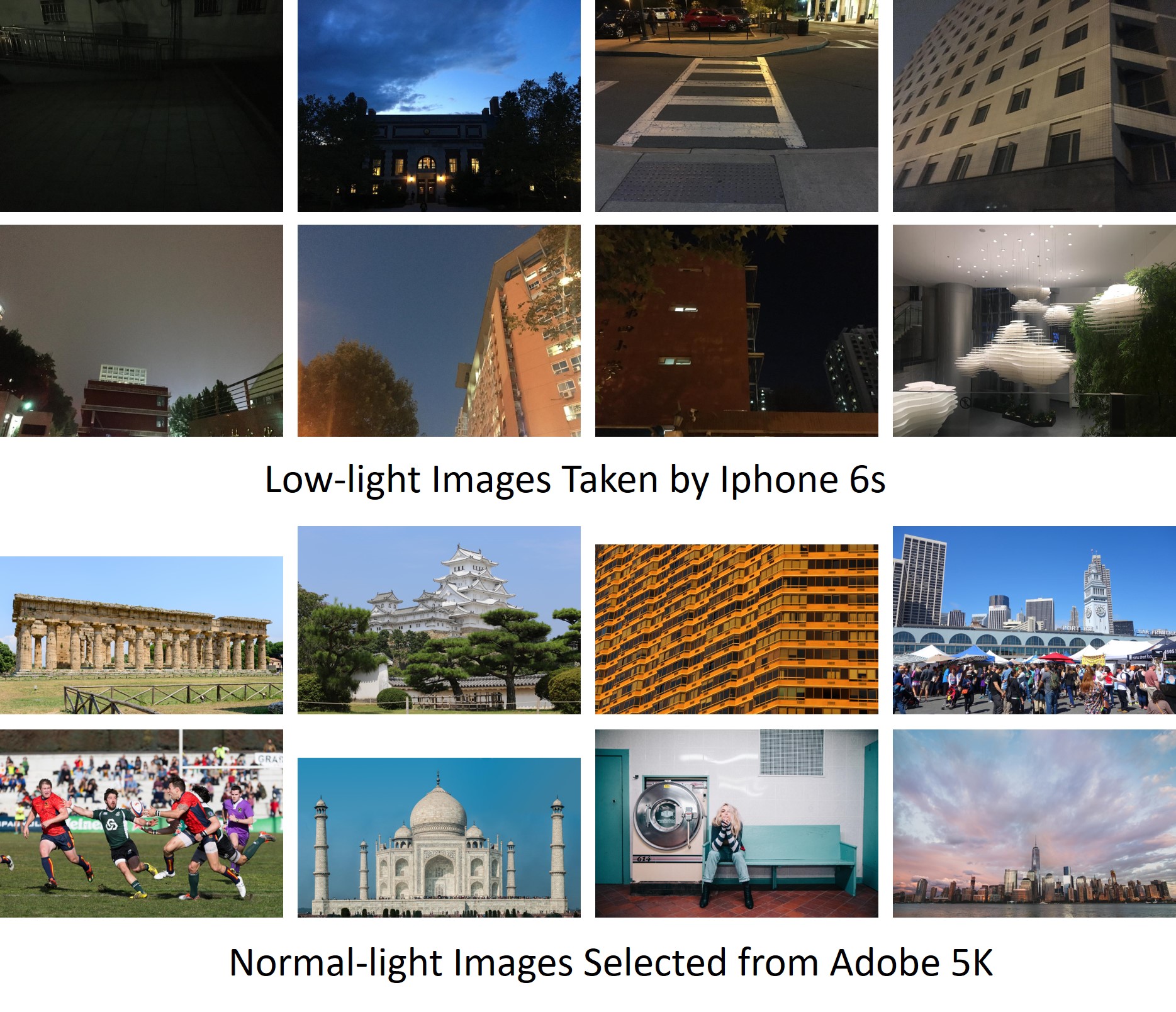}
    \caption{Examples from our URL dataset. The top two rows display the low-light images collected with an iPhone 6s cell phone. We can see that the illumination conditions are quite diverse. Some images are very dark. The bottom two rows display the normal-light natural images collected from Adobe 5K. }
    \label{fig:train_examples}
\end{figure}

\begin{figure*}
    \centering
    \includegraphics[width=\textwidth]{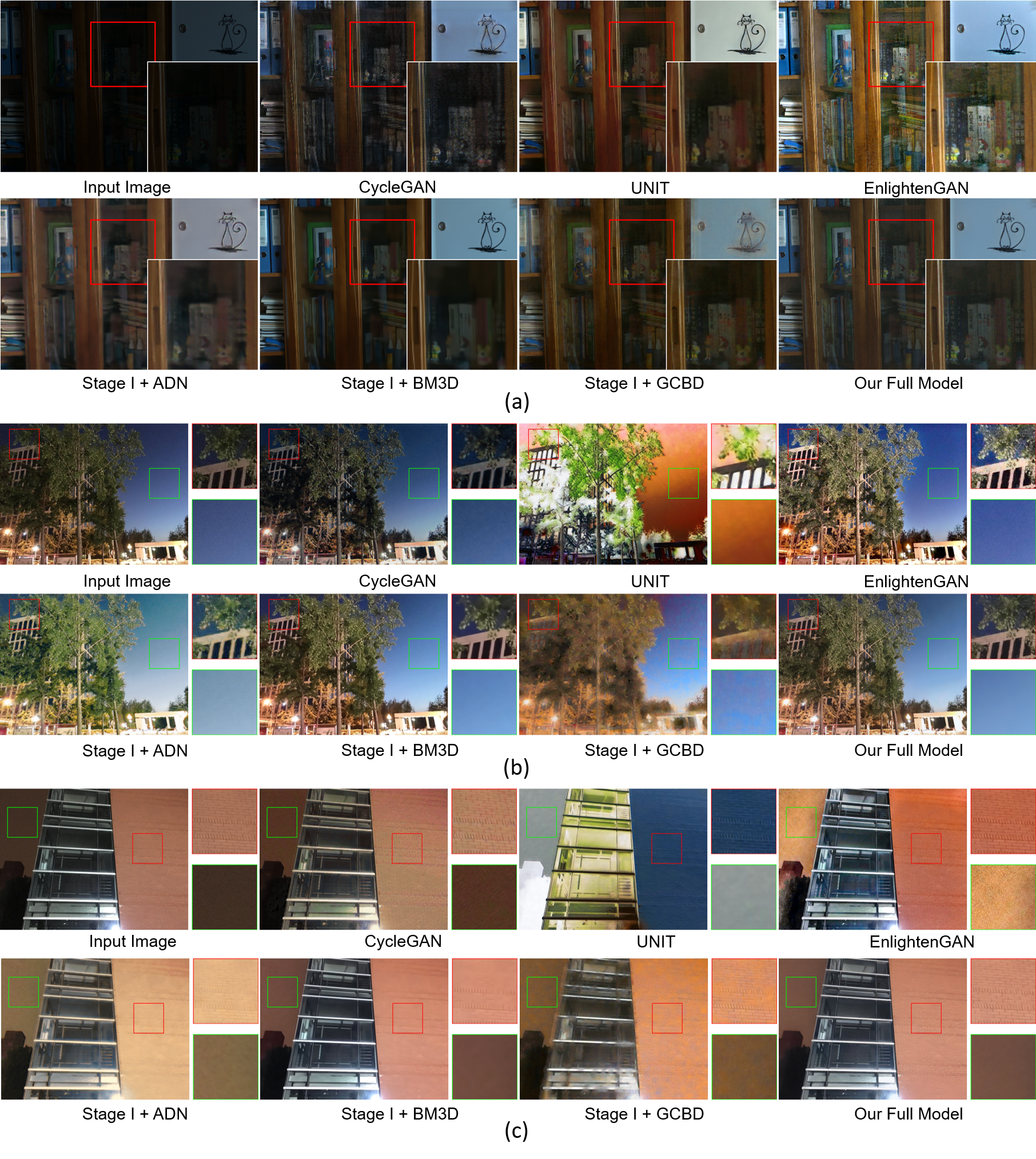}
    \caption{Qualitative results (enhancement with denoising) on the LOL  (a) and URL  (b and c) datasets. \textit{Please zoom in to the maximum for the very details}.}
    \label{fig:lol}
\end{figure*}

\section{Experiments}
In this section, we first compare the performance of each model with respect to only illumination enhancement on an unpaired enhancement dataset from EnlightenGAN \cite{jiang2019enlightengan} which \textit{does not contain noticeable noise}. Then we conduct experiments for both illumination enhancement and noise suppression on LOL dataset \cite{wei2018deep} and our collected unsupervised real-world low-light dataset (URL dataset), which \textit{both contain low-light images with noticeable noise}.

\vspace{-0.1in}
\subsection{Datasets}

\textbf{Unpaired Enhancement Dataset}. Jiang et al. \cite{jiang2019enlightengan} collect an unpaired dataset for training contrast enhancement models. The training set is composed of 914 low-light images which are dark yet \textit{containing no significant noise}, and 1016 normal-light images from public datasets. We use this dataset to compare the performance of contrast enhancement of each model.  The evaluation set is composed of 148 low-light/normal-light image pairs from public datasets. All the images from both the training and evaluation sets have been resized to $400\times600$. 

\textbf{LOw-Light (LOL) Dataset} \cite{wei2018deep}. LOL is composed of 500 low-light and normal-light image pairs and is split into 485 training pairs and 15 testing pairs. The low-light images contain noise produced during the photo capture process. Most of the images are indoor scenes. To adapt the dataset to our unsupervised setting, we adopt the training images as our low-light train set and adopt the normal-light images in the Unpaired Enhancement Dataset \cite{jiang2019enlightengan} as the normal-light train set. The testing images remain the same as the LOL dataset. All the images have a resolution of $400\times600$.

\textbf{URL Dataset}. There are a quite limited number of real-world low-light datasets publicly available. Among the public low-light datasets, some of them are composed of synthetic images while many other datasets such as ExDark \cite{loh2019getting} or Adobe FiveK \cite{fivek} contain dark images without significant noise. These datasets do not meet the objective of our study. Therefore, we collect an Unsupervised Real-world Low-light dataset (URL dataset) composed of $414$ high-resolution real-world low-light images taken by iPhone-6s and $3,837$ normal-light images selected from Adobe FiveK. To collect the low-light images, we first take photos with an iPhone-6s from various scenes in different cities around the world. We remove images that are too dark (cannot be recovered since details are lost), blurry, or with high brightness. We also remove images that are very similar to  other images in order to boost the diversity of the dataset. In the end, we are able to select 414 low-light images from over 4,000 photos. Our URL dataset is quite diverse, containing various indoor and outdoor scenes under different light conditions. Consequently, the level of noise contained in each image or even different regions of the same image varies considerably across the dataset. We divide the low-light images into 328 training images and 86 testing images. This dataset thus complements the existing datasets in those two regards.  Note that there is no corresponding ground-truth image for each testing image. Each low-light image is resized to $1008\times756$. Fig. \ref{fig:train_examples} shows low-light image examples from our dataset.

\begin{table}[t]
	\centering
	
	\caption{ Quantitative results for illumination enhancement on the Unpaired Enhancement Dataset.}
	
	\begin{tabular}{l|cccc}
		\hline
		Model & CycleGAN  & UNIT & EnlightenGAN & Ours Stage I \\
		\hline
		\hline
		PSNR &18.22 &8.42 &17.31 & \textbf{19.78 }\\
		\hline
		SSIM &0.7284 &0.2549 &0.8047 &\textbf{0.8197} \\
		\hline
	\end{tabular}
	\label{table:contrast_quantitative}
\end{table}

\subsection{Implementation Details}
\textbf{Implementation Details of Stage I}
In Stage I, we first train the networks with a learning rate of $0.0001$ using Adam optimizer for 100 epochs, then the learning rate is decreased linearly to zero within the next 100 epochs. We split the training data into train and validation sets, and select the best model on the validation set, then apply the model to the test images. We use a batch size of 32. We crop $320\times320$ patches for training and randomly flip the patches for data augmentation. The whole model is trained on two 1080Ti GPUs. 

\textbf{Implementation Details of Stage II}
In Stage II, we train the networks with an initial learning rate of $0.0001$ for $2,000$ epochs. We train the model with randomly cropped image patches (with a size of $128\times128$), as our model in this stage primarily needs to capture the noise pattern in the local regions. We randomly rotate and flip the patches for data augmentation. The whole model is trained on two 1080Ti GPUs.

\subsection{Experiments for Illumination Enhancement }
We compare our Stage I model which does only illumination enhancement on the Unpaired Enhancement Dataset \cite{jiang2019enlightengan} with state-of-the-art models, including CycleGAN \cite{zhu2017unpaired}, UNIT \cite{liu2017unsupervised} and EnlightenGAN \cite{jiang2019enlightengan}. As shown in Fig. \ref{fig:contrast}, our model can generate normal-light images with reasonable contrast and color in both global and local regions. EnlightenGAN can produce visually pleasing images. However, they may still suffer from color distortion in several local regions, as indicated by the red boxes. The other unsupervised image translation methods can synthesis roughly good images. However, the color, contrast are not perfect. 

We report the PSNR and SSIM of the generated images \textit{as a complement} to the visual results on Unpaired Enhancement Dataset.  Results in Table \ref{table:contrast_quantitative} show that our model performs significantly  better than the existing models, which are consistent with the visual results.

\begin{figure*}[t]
    \centering
    \includegraphics[width=\textwidth]{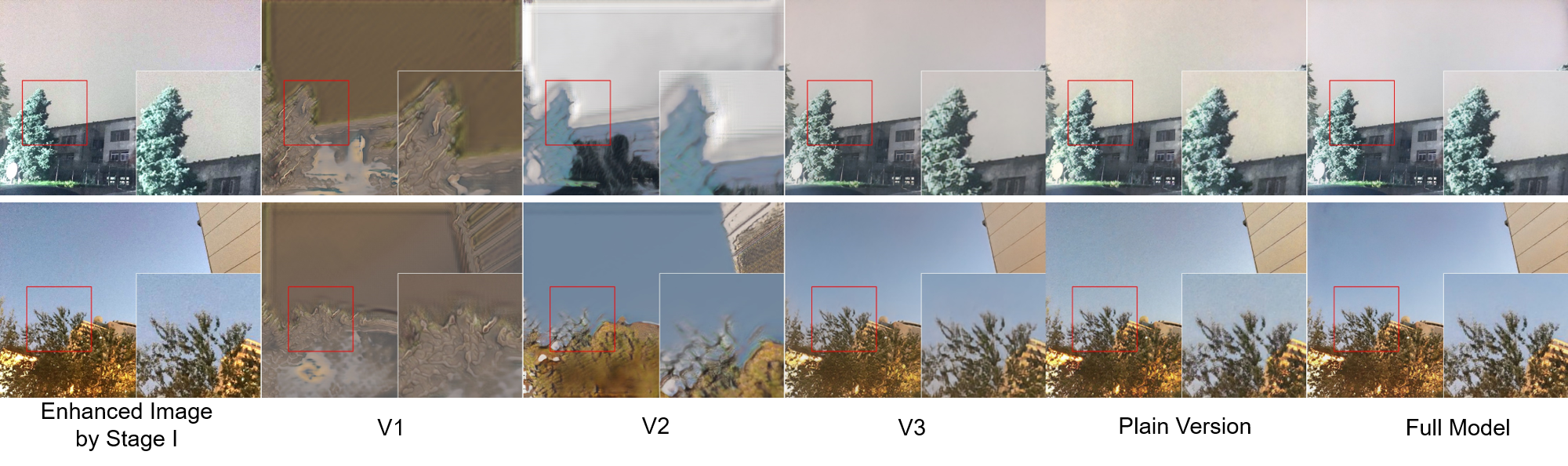}
    \caption{Ablation Study on our URL dataset. We strongly encourage readers to zoom in to the maximum for the very details.}
    \label{fig:ablation_url}
\end{figure*}

\subsection{Experiments for Both Illumination  Enhancement and Noise Suppression}
\textbf{Experiment Settings.}
We train the two stages of our model separately, as we observe that jointly training the two stages yields unstable results under our unsupervised learning setting. 
We evaluate our model on the real-world low-light image enhancement datasets: LOL and our URL datasets, and compare it with the state-of-the-art unsupervised image-to-image translation or contrast enhancement models, including CycleGAN \cite{zhu2017unpaired}, UNIT \cite{liu2017unsupervised} and EnlightenGAN \cite{jiang2019enlightengan}. We also compare our model with the combination of illumination enhancement model and denoising model. Specifically, we compare our full model with our Stage I + BM3D \cite{dabov2007image}, our Stage I + ADN \cite{adn2019_miccai,adn2019_tmi} and our Stage I + GCBD \cite{chen2018image}. BM3D \cite{dabov2007image} is a robust image denoising method. The limitation is that it requires a known noise level as input. To use BM3D in our task, we first estimate a rough noise level for each testing image, then apply BM3D on the testing images.

\begin{table}[t]
	\centering
	
	\caption{Quantitative results on the LOL dataset.}
	
	\begin{tabular}{l|c|c|c}
	    \hline
	    Dataset &\multicolumn{3}{c}{LOL} \\
		\hline
		Model & PSNR $\uparrow$ & SSIM $\uparrow$ & LPIPS $\downarrow$  \\
		\hline
		\hline
		CycleGAN &14.75 &0.6852 & 0.3808\\
		UNIT     &15.49 &0.7280 & 0.3476\\ 
		EnlightenGAN &18.36 &0.7839 & 0.2915\\
		Stage I + ADN &17.72  &0.7776 & 0.3531\\
		Stage I + BM3D &19.36 &0.8154 & 0.2837\\
		Stage I + GCBD & 18.76 & 0.7753 & 0.3579\\
		Our Full Model &\textbf{20.04} &\textbf{0.8216} & \textbf{0.2661}\\
		\hline
	\end{tabular}
	\label{table:denoise_quantitative}
\end{table}

\begin{table}[t]
	\centering
	
	\caption{User Study on the URL dataset.}
	
	\begin{tabular}{l|c}
		\hline
		Model &  POS \\
		\hline
		\hline
		CycleGAN &3 \\
		UNIT     &3\\ 
		EnlightenGAN &28 \\
		Stage I + ADN &15 \\
		Stage I + BM3D &64 \\
		Stage I + GCBD  &2 \\
		Our Full Model &\textbf{85} \\
		\hline
	\end{tabular}
	\label{table:userstudy}
	
\end{table}

\textbf{Quantitative Results. }
On the LOL dataset, we report PSNR, SSIM results. We also report the perceptual score (LPIPS) of the enhanced images to better quantify the perceptual quality of images. From Table~\ref{table:denoise_quantitative} we can see that our model is consistently better than the existing models, demonstrating the superiority of our decoupled networks.  

\textbf{Qualitative Results. }
Fig. \ref{fig:lol} show the qualitative results on both the LOL dataset and our URL dataset. CycleGAN generates heavy artifacts and slightly suffers from color distortion. UNIT suffers from heavy color distortion on the URL dataset and cannot preserve details on the LOL dataset.  EnlightenGAN can improve the illumination of the images, but there are still obvious noise and artifacts on the image.  \textit{ The visual results indicate that it is challenging to handle image illumination enhancement and denoising simultaneously with a single model for real-world low-light image enhancement.}

However, simply cascading an illumination enhancement model with a denoising model still cannot produce satisfactory results. From Fig.~\ref{fig:lol}, we see that the results of ADN and GCBD still exhibit color distortion or contain many artifacts on both datasets. ADN even suffers from overexposure in bright regions on URL dataset. Using BM3D to post-process the results of our Stage I yields good color and illumination. However, from Fig.~\ref{fig:lol}, we observe that the texture details of images on both the LOL and URL datasets are over-smoothed. A possible reason is that BM3D was proposed for synthetic noise removal. It may not generalize well to real-world denoising. Compared to existing methods, our full model performs noise removal adaptively and considers preserving the details and color when denoising. Therefore, It can suppress noise as well as preserving more details.

\textbf{User Study.} 
We perform a user study on the URL dataset. Specifically, we randomly select 20 low-light testing images from the URL dataset. For each input image, we show each user the final enhanced result of each method and ask the user to select the most visually pleasing result among all methods, considering illumination condition, color, texture realness, and noise level. In total, we obtain 200 preference opinions.  Table \ref{table:userstudy} shows the preference opinion score (POS) of each method. Our method outperforms state-of-the-art methods. It is worth noting that although EnlightenGAN can generate images with good illumination, it fails to remove noise when noise is heavy, as shown in Fig. \ref{fig:lol}. As a result, its POS is relatively low. The other one-stage models CycleGAN and UNIT suffer from color distortion. The results further indicate that it is challenging for a single model to perform illumination enhancement and noise removal simultaneously. 

\begin{table}[t]
    \centering
	\setlength\tabcolsep{0.2cm}
	\caption{ Configuration of different versions of our model.}
	
	\begin{tabular}{l|c|c|c|c}
		\hline
		Model & $L_G$ & $L_{color}$ & $L_{con}^{adapt}$ & $L_{con}$  \\
		\hline
		\hline
		Version 1 & \checkmark & & & \\
		Version 2 &\checkmark &\checkmark & & \\
		Version 3 &\checkmark &\checkmark &\checkmark &\\ 
		Full Model &\checkmark &\checkmark &\checkmark &\checkmark\\		Plain &\checkmark &\checkmark & vanilla & \checkmark \\
		\hline
	\end{tabular}
	\label{table:ablation_list}
\end{table}

\begin{table}[t]
    	\setlength\tabcolsep{0.3cm}
    	\centering
	\caption{Non-ref image quality scores on URL. (lower is better)}
	\begin{tabular}{l|c|c|c}
		\hline
		Model & V3 & Plain & Full  \\
		\hline
		\hline
		Brisque &33.66 &26.95 &\textbf{22.13} \\
		\hline
		NIQE &4.12 &3.44 &\textbf{2.53} \\
		\hline
	\end{tabular}
	\label{table:ablation}
\end{table}

\begin{figure*}[t]
    \centering
    \includegraphics[width=0.75\textwidth]{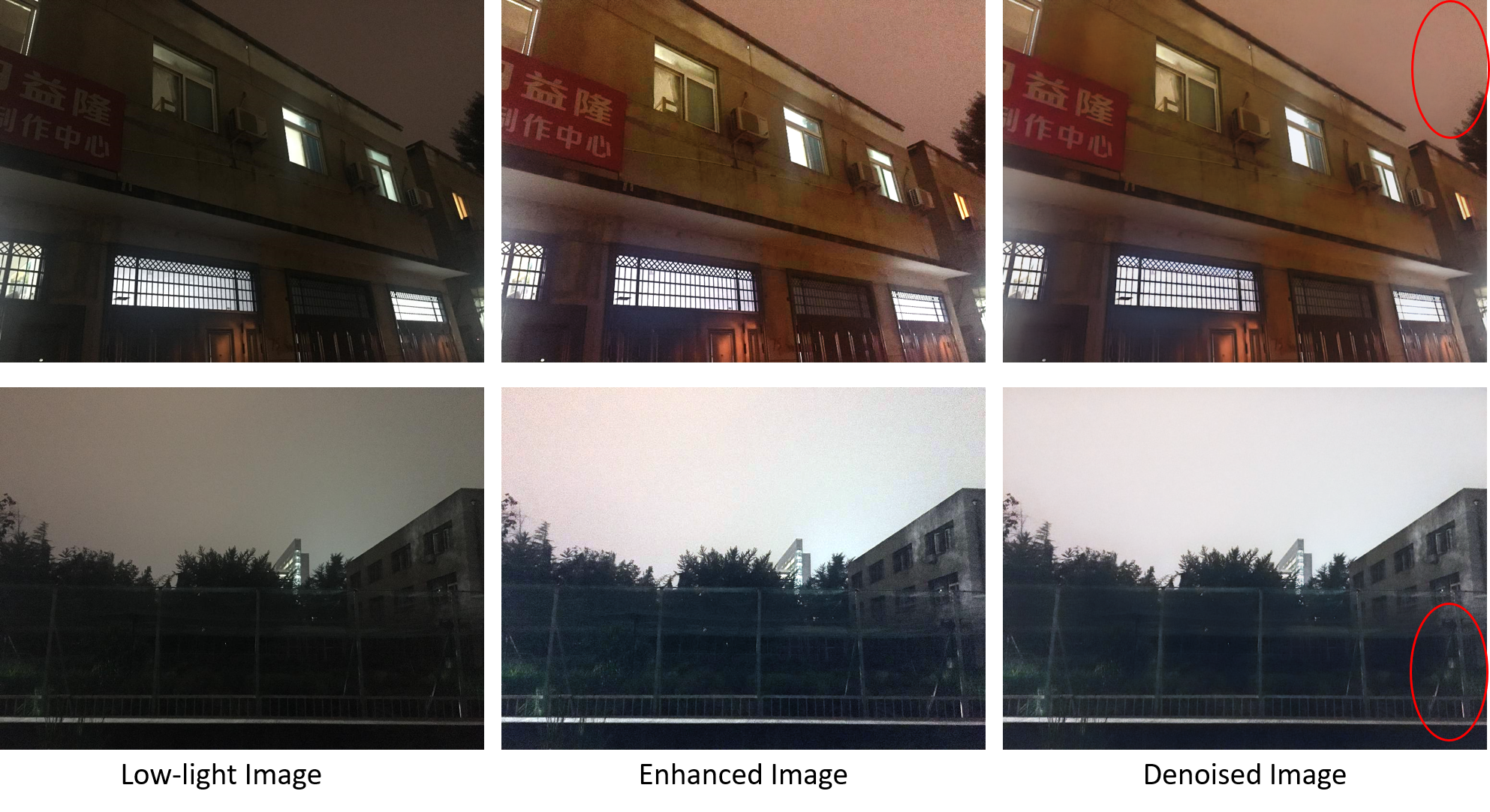}
    \caption{Failure cases of our method. In each row, from left to right we present the input low-light image, the enhanced image from our Stage I model, and the final denoised result from our Stage II model. Please zoom in to see the details and pay special attention to the red marked regions.}
    \label{fig:failure}
\end{figure*}

\subsection{Ablation Study}
In this section, we study how each component of our model contributes to the final performance. We primarily analyze the components in Stage II, which are the core contributions in this work. Specifically, as shown in Table~\ref{table:ablation_list}, we compare the versions of our Stage II model with different losses imposed. Besides the versions regarding the loss functions, we also compare our model to a non-adaptive denoising model, which we call the Plain Model, \textit{i.e.}, the generator only takes the enhanced image as input without illumination guidance. The four losses used in the Plain Model remain the same as our Full Model, except that we use the vanilla content loss instead of adaptive content loss. 

Experiments are conducted on our URL dataset. From Fig. \ref{fig:ablation_url} we observe that the results from Stage I still contain heavy noise, further indicating that {it is challenging to model both illumination and noise patterns with a single network}.  Merely using adversarial loss (Version 1) can help to smooth the image, but the color is shifting and cannot be well controlled. Using the color loss (Version 2) significantly helps to constrain the output image to have similar color and contrast as the input image. However, without learning with \textit{pseudo triples} and adaptive content loss (i.e., loss $L_{con}^{adapt}$), the content of the output is not explicitly constrained and the images contain distorted contextual details.  When imposing the learning with \textit{pseudo triples} (Version 3), the network can  produce realistic contents and perform noise suppression, \textit{indicating that Pseudo Labeling plays a key role in stabilizing the training and improving the performance of noise modeling}. However, several local regions still lack details. Please pay attention to the trees in these results in Fig. \ref{fig:ablation_url}. The texture details of the trees and leaves are smoothed in Version 3 and preserved in our Full Model, indicating that the content loss between the real input image and its output image further helps to preserve contextual details during denoising. 

Comparing our Full Model in Stage II with the Plain Model, we observe that the Plain Model cannot effectively suppress the noise. There is still notable noise in the sky region and other regions, as shown in Fig. \ref{fig:ablation_url}. A possible reason is that the plain model may not be able to perceive all the noise patterns under different illumination conditions precisely. In contrast, our model is explicitly guided by the illumination of the image. Therefore it can capture the noise pattern under various illumination conditions more effectively and produce better results.  

Since the URL dataset does not contain ground truth images, we use non-reference image quality assessment methods BRISQUE \cite{mittal2012no} and NIQE \cite{mittal2012making} to quantify the quality of the testing images. Results are shown in Table \ref{table:ablation}. We do not include the scores of V1 or V2, as these two versions can only produce heavily distorted images, as shown in Fig. \ref{fig:ablation_url}. These quantitative results further demonstrate the importance of using illumination as guidance for real-world noise modeling.

\subsection{Failure Cases}
Fig. \ref{fig:failure} shows two failure cases of our model on the URL dataset. In the top case, there is still noise at the upper-right corner of the denoised image. In the bottom case, there is still noise at the bottom-right corner of the denoised image, as indicated by the red circles. These results demonstrate our model's limitation on dealing with image borders or corners. A possible reason may be that the noise pattern at the corners or the borders of the image may be different from that of the other regions, since we are dealing with the real-world noise that is spatially variant. We will continue working on it to improve our model in the future. 

\section{Conclusion}
In this paper, we have presented the decoupled networks to address the real-world low-light image enhancement problem in an unsupervised fashion. 
Our model in Stage I enhances a low-light image to generate an image with satisfactory illumination and color. Our model in Stage II further denoises the enhanced image to obtain a clean image, while preserving good contrast, color and contextual details. We conduct experiments on three real-world datasets. The results show that our model outperforms the state-of-the-art models in terms of both illumination enhancement and noise removal.

\appendix[Architecture of our model]
\subsection{Generator of Stage I}
The generator of Stage I is composed of an encoder, a pyramid module, and a decoder. The encoder is composed of several convolutional blocks and Max pooling layers. Each convolutional block is composed of a convolution layer, a Batch Normalization layer (except the first convolutional block), and a Leaky ReLU activation function with slope to be $0.2$.  

The decoder is composed of three Deconvolutional blocks. Each Deconvolutional block is composed of an upsampling layer, two convolution layers, a Batch Normalization layer and a Leaky ReLU activation layer (the activation of the last Deconvolutional block is Sigmoid instead of Leaky ReLU ).

The overall structure of our generator in Stage I is: $CCMCCMCCMPDDD$, where $C$ denotes to Convolutional block, $M$ means Max pooling, $P$ means pyramid module, $D$ means deconvolutional block. 

\subsection{Discriminators of Stage I}
In Stage I, we have a global discriminator and a local discriminator. The global discriminator is composed of five convolutional layers. Each convolution layer is followed by a Leaky ReLU activation function layer, except the last layer. The local discriminator has the same structure as the global discriminator. 

\subsection{Generator of Stage II}
The generator of Stage II is composed of an encoder, six Resnet blocks and a decoder. The encoder is composed of two convolutional blocks. The decoder is composed of three deconvolutional blocks, as introduced in the previous section. The detailed structure of our generate in Stage II is: $CCRRRRRRDDD$, where $R$ means a Resnet block. 

\subsection{Discriminator of Stage II}
We use two-scale discriminators in Stage II. Each discriminator has the same structure as the discriminators used in Stage I. The only difference between discriminators at different scale is the input. The input to the first discriminator is the original full-sized image. The input to the second discriminator is a downsized version of the full-sized image. We use Average pooling to downsize the image. 


%

\ifCLASSOPTIONcaptionsoff
  \newpage
\fi



%

\bibliographystyle{IEEEtran}
\bibliography{IEEEfull}

\end{document}